\documentclass[
aps,nofootinbib,
showpacs,showkeys,preprint
tightenlines,preprintnumbers,] {revtex4}

\usepackage{epsf,epsfig,subfigure,graphicx,amsmath,amssymb 
}
\usepackage{color}
\newcommand{\dis}[1]{\begin{equation}\begin{split}#1\end{split}\end{equation}}

\newcommand{\etal}{{\it et al.\,}}

\newcommand{\Qem}{Q_{\rm em}}

\newcommand{\gev}{\,\textrm{GeV}}

\newcommand{\eV}{\,\mathrm{eV}}

\def\sw0{{$\sin^2\theta_W^0$}}

\def\E6{{\rm E_6}}

\def\EE8{{\rm E_8\times E_8'}}

\def\three{{\bf 3}}

\def\one{{\bf 1}}

\def\two{{\bf 2}}

\def\three{{\bf 3}}

\begin{document}

\draft

\title{\Large\bf   Anticorrelation of Mass and Mixing Angle Hierarchies
}

\author{Paul H. Frampton$^{(1)}$ and Jihn E.  Kim$^{(2,3)}$ }

\address
{$^{(1)}$ Dipartimento di Matematica e Fisica ``Ennio De Giorgi'',
Universita del Salento and INFN-Lecce,
Via Arnesano, 
73100 Lecce, Italy  \\
$^{(2)}$Department of Physics, Kyung Hee University, 26 Gyungheedaero, Dongdaemun-Gu, Seoul 02447, Republic of Korea  \\
 $^{(3)}$ Department of Physics and Astronomy, Seoul National University, 1 Gwanakro, Gwanak-Gu, Seoul 08826, Republic of Korea
}

\begin{abstract} 
We obtain  a relationship between the hierarchies of mixing 
angles and of masses pertinent
to the Cabibbo-Kobayashi-Maskawa (CKM) 
quark mixing matrix and the Pontecorvo-Maki-Nakagawa-Sakata (PMNS)
lepton mixing matrix. Using this relationship, we argue that the more severe hierarchy of the
charge-$\frac{2}{3}$ quark masses requires that the
CKM matrix be close to a unit matrix whereas the
milder hierarchy of the neutrino masses allows the PMNS 
matrix to depart markedly
from the CKM matrix and contain large mixing angles of the 
type that are observed. 
\end{abstract}
\maketitle


\section{Introduction}\label{sec:Introduction}

The Standard Model (SM) of particle physics has many free parameters which are fitted to experimental data rather than calculated.  Most of these parameters are masses and mixing angles, and in the present article we demonstrate a general model-independent relationship.  Understanding these from symmetric principles is a theoretical solution of the flavor problem.
On the other hand, a generally accepted idea in the astrophysical
community is most of the energies in the universe are carried by 
dark energy (DE) and dark matter (DM).
The solution of the flavor problem is applicable to the cosmic evolution because the symmetry required from the flavor solution necessarily dictates the
properties of the beyond-the-SM (BSM) particles, in particular 
DM. Not only DM but also the baryon number in the universe (BAU) is related to the theoretical solution because one of three Sakharov conditions \cite{Sakharov67} for generating the BAU needs C and CP violation. The CP phase in the SM may be directly related to
 the BAU \cite{Type2Leptogenesis} or may not be related
 \cite{Segre80,Yana86}. Even though the CP phase needed for the BAU may not be that of the SM,  the phases of the  SM and the BSM can be related if the symmetry in the full theory is known.

Among the SM parameters, the best measured parameters are
the gauge coupling constants and some masses  $m_e, m_\mu,m_t$  and $m_h$. Other masses have rather large error bars. 
 On the other hand, three real angles of the Cabibbo-Kobayashi-Maskawa (CKM)  matrix and three
real angles of the  Pontecorvo-Maki-Nakagawa-Sakata (PMNS) matrix are known rather accurately. The CP phases are not known accurately.  Therefore, the flavor symmetry can be grasped by looking at the CKM and the PMNS
matrices with the phases as free parameters \cite{Kimetal19}.
The CKM matrix is close to identity, but the PMNS matrix is of the 
form
 \dis{
 V^{(0)}\sim \begin{pmatrix}
 \times & \times &  \times \\
 \times & \times & \times \\
0 & \times & \times
 \end{pmatrix},\label{eq:PMNSform}
 }
where $\times$ are non-vanishing entries. The form of Eq. (\ref{eq:PMNSform}) has led to the bimaximal 
PMNS form \cite{Perkins95}. It is almost   tri-bimaximal  \cite{NuFIT18}, which created a frenzy with regard to the tetrahedral symmetry \cite{Ma02}.

\section{Form of charged currents}

At present all the SM particles of three quark-lepton families are known, and one of the most important 
theoretical problems is the flavor problem. The first SM family consists of fifteen chiral fields, with three colors $(\alpha=1,2,3)$ for the quarks,
\dis{
{Q}^{\alpha }_{L} ,~ (u^\alpha)^c_L,~(d^\alpha)^c_L,~{\ell}_{L} ,~e^c_L,
}
where
\dis{
{Q}^{\alpha}_{L} =\begin{pmatrix} u^{\alpha}\\[0.3em] d^{\alpha }\end{pmatrix}_L,~{\ell}_{L} =\begin{pmatrix} \nu_e \\[0.3em] e\end{pmatrix}_L,
}
and the 2nd and the 3rd families have an identical structure except for the masses and mixings, which
are our present research topic. Except for the three neutrinos, all the SM chiral fields obtain masses by using renormalizable couplings 
to the vacuum expectation value (VEV) of a Higgs doublet. Neutrinos remain massless unless additional assumptions beyond the 
Standard Model (BSM) are invoked.

We have the irreducible representations of the gauge group $SU(3)_C \times SU(2)_L \times U(1)_Y$ that we put into the definition 
of Yukawa couplings. Let us choose the bases where  
the fermion doublets with the 3rd component of the weak isospin  $T_3=-\frac12$ are the mass eigenstates:
\dis{
d^{(\rm mass)}_{L} =\begin{pmatrix} d^{(0)} \\[0.3em] s^{(0)}\\[0.3em] b^{(0)}\end{pmatrix}_L=
\begin{pmatrix} d \\[0.3em] s\\[0.3em] b\end{pmatrix}_L,
~ e^{(\rm mass)}_{L} =\begin{pmatrix} e^{(0)} \\[0.3em] \mu^{(0)}\\[0.3em] \tau^{(0)}\end{pmatrix}_L 
=\begin{pmatrix} e \\[0.3em] \mu\\[0.3em] \tau\end{pmatrix}_L.
}
Then, the charged currents (CCs) are given by 
\dis{ 
 \frac{g}{\sqrt2}\left( \bar{u}^{(\rm 0)}_{L} \gamma^\mu d^{(\rm mass)}_{L}  +   \bar{\nu}^{(\rm 0)}_{L} 
 \gamma^\mu e^{(\rm mass)}_{L}\right) W^+_\mu,
}
where
\dis{
{u}^{(\rm 0)}_{L} =\begin{pmatrix} u^{(0)}\\[0.3em] c^{(0)}\\[0.3em] t^{(0)}\end{pmatrix}_L,~  {\nu}^{(\rm 0)}_{L} 
=\begin{pmatrix} \nu_e^{(0)}\\[0.3em] \nu_\mu^{(0)}\\[0.3em] \nu_\tau^{(0)}\end{pmatrix}_L.
}
In terms of mass eigenstates,  $\Qem=+\frac23$ quarks ($u,c,t$) and  neutrinos  ($\nu_1,\nu_2,\nu_3$), the defining 
weak states are related to the mass eigenstates by
\dis{
\begin{pmatrix} u^{(0)}\\[0.3em] c^{(0)}\\[0.3em] t^{(0)}\end{pmatrix}_L= U^{(u)\,\dagger}
\begin{pmatrix} u \\[0.3em] c \\[0.3em] t \end{pmatrix}_L,~~  \begin{pmatrix} \nu_e^{(0)}\\[0.3em] \nu_\mu^{(0)}\\[0.3em] 
\nu_\tau^{(0)}\end{pmatrix}_L= U^{(\nu)\,\dagger}\begin{pmatrix} \nu_1\\[0.3em] \nu_2\\[0.3em] \nu_3\end{pmatrix}_L,
}
where the unitary matrices diagonalizing L-handed fields are denoted by $U$ and the unitary matrices diagonalizing
 R-handed fields are denoted by ${\cal U}$. 
Then, the CKM and the PMNS matrices are given by
\dis{
V^{\rm (CKM)}= U^{(u)} U^{(d)\,\dagger}= U^{(u)} ,~V^{\rm (PMNS)}=U^{(\nu)} U^{(e)\,\dagger}= U^{(\nu)} .\label{eq:VandU}
}
The definitions of $ U^{(u)}$ and $U^{(\nu)}$ in Eq. (\ref{eq:VandU}) have the required number of parameters. In  $ U^{(u)}$,  two phases of L-handed $ u^{(0)}$ quarks for constraints exist because the baryon number phase cannot be 
used as a constraint. Also, three  $ u^{(0)}$ masses provide three constraints. Thus, out of 9 parameters in a $3\times 3$ 
unitary matrix, the number of undetermined parameters are 4: 3 real angles and 1 phase. In  $ U^{(\nu)}$, we do not have 
any phase constraint because Majorana neutrinos are real,  we have nine minus three parameters: 3 real angles, 1 Dirac 
phase, and 2 Majorana phases.

Note that our definition of the PMNS matrix is given in accordance with the CKM matrix by using the $W^+_\mu$ coupling
while the Particle Data Book (PDG) defines the PMNS matrix with the $W^-_\mu$ coupling \cite{PDG18PMNS,PDG18}. This is because 
the quark mixing angles  are represented with respect  to the mass eigenstates while the leptonic mixing angles  are
 represented with respect  to the weak eigenstates.  Thus, what the PDG book  represents is our $U^{(\nu)\dagger}$.
Therefore, in discussing the CKM and the PMNS matrices in unison, considering the charge raising currents 
due to $U^{(u)}$ and  $U^{(\nu)}$ to be the same is better.
We could have chosen the mass eigenstates of $\Qem=+\frac23$ quarks and neutrinos as the defining fermions with 
$^{(0)}$, but then we would have to specify how neutrinos obtain masses, which necessarily includes BSM physics. 
On the other hand, our choice of  $\Qem=-\frac13$ quarks and   $\Qem=-1$ leptons as the defining fermions is fulfilled 
just by the above fermions and the Higgs doublet $H_d$ with $Y=-\frac12$:
\dis{
  \left[ \overline{q}_{\alpha L}\begin{pmatrix}m_d&,0,&0\\
   0,&m_s,&0\\
   0,&0,&m_b \end{pmatrix} d_R^{\alpha \rm (mass)}+ \overline{\ell} \begin{pmatrix}m_e&,0,&0\\
   0,&m_\mu,&0\\
   0,&0,&m_\tau \end{pmatrix} e_R^{\rm (mass)} \right] \frac{H_d}{\langle H_d^0\rangle}+{\rm h.c.}\label{eq:DownTypeY}
}
where ${q}_{i L}=(u^{(0)},d^{\rm (mass)})^T_{iL}  $.
This is equivalent to defining the Yukawa couplings as $Y^{(d)}_{ij}= \frac{m^{(d)}_i}{\langle H_d^0\rangle}\delta_{ij}$ 
and $Y^{(e)}_{ij}= \frac{m^{(e)}_i}{\langle H_d^0\rangle}\delta_{ij}$.
The renormalizable couplings for the   $\Qem=+\frac23$ quarks are given by $H_d^\dagger$ that carries  $Y=+\frac12$.

\bigskip

\noindent
For neutrinos, renormalizable couplings such as  Eqs. (\ref{eq:DownTypeY})  
cannot be written.  If non-renormalizable 
couplings are allowed, however, lepton (L) number can be violated and L-violating Majorana-type neutrino mass terms are possible \cite{Weinberg79}:
\dis{
\frac{k_{ij}}{M}\,(\ell_i^T)^{m} C^{-1}(\ell_j)^{n}\, (H_u)^{m'} (H_u)^{n'}\epsilon_{mm'}\epsilon_{n n'}, \label{eq:NuTypeY}
}
where $m , n ,m'$ and $n'$ are the SU(2) indices and $i$ and $j$ are flavor indices. For a non-vanishing VEV of 
$H_u$, $\langle H_u^0\rangle=v_u/\sqrt2$,  which has $T_3=-\frac12$; neutrinos having $T_3=+\frac12$ lead to the mass
\dis{
m_{ij}= \frac{v_u^2 }{M} k_{ij} .
}

\section{Small quark mixing and large lepton mixing angles}

\vskip 0.5cm
\noindent{\bf 1. The CKM Matrix}\vskip 0.5cm 

The CKM matrix elements are
\dis{
{\rm CKM}~({\rm component}\,ij):~&\frac{g}{\sqrt2}\, \bar{u}_{iL}V^{\rm (CKM)}_{ij} \, \gamma^\mu   \, {d} _{jL}\, W^+_\mu\,,
}
where $u_{(i)}$ and $d_{(i)}$ are mass eigenstates.
 The diagonal masses of $\Qem=+\frac23$ quarks are, for the  $(ll)$ component, 
\dis{
{\rm Diagonal}~(ll):&~ m_u\, \bar{u}_{R}u_{L}+m_c\, \bar{c}_{R}c_{L}+m_t\, \bar{t}_{R}t_{L}+{\rm h.c.}\\
&=m_u\, \bar{u}^{(0)}_{iR}\,({\cal U}^\dagger_{i1}U_{1j})u^{(0)}_{jL}+m_c\, \bar{u}^{(0)}_{iR}\,
({\cal U}^\dagger_{i2}U_{2j})u^{(0)}_{jL}+m_t\, \bar{u}^{(0)}_{iR}\,({\cal U}^\dagger_{i3}U_{3j})u^{(0)}_{jL}+{\rm h.c.}\\
& =\bar{u}^{(0)}_{\alpha R}\,(  m_i {\cal U}_{i \alpha}U_{i\beta})u^{(0)}_{\beta L}+{\rm h.c.}\label{eq:TopBasic}
}
Let us pay attention to the largest term
\dis{
& m_{t}\,(\bar{u}^{(0)}_{\alpha R} {\cal U}_{t\alpha }U_{t\beta}u^{(0)}_{\beta L}+  O(\frac{m_c}{m_t}) )
= m_{t}\,(\bar{u}^{(0)}_{\alpha R} {\cal  U}^\dagger _{\alpha t})(U_{t\beta}u^{(0)}_{\beta L})+  O(m_c).  
}
Thus, we define
\dis{
t_L\simeq U_{t\alpha}u^{(0)}_{\alpha L}\,,\\
t_R\simeq {\cal U}_{t\beta}u^{(0)}_{\beta R}\,.\label{eq:HeavyTop}
}
Whatever the diagonalizing matrices $U$ and ${\cal U}$ may be, we can approximately find the top 
components by using Eq. (\ref{eq:HeavyTop}). With the choice of $t_L$ and $t_R$,   the mass matrix is of a form
with a determinant $O(\epsilon^3)$:
\begin{equation}
m_t\begin{pmatrix}
O(\varepsilon^2) &O(\varepsilon^{\frac{3}{2}})) &O(\varepsilon) \\
O(\varepsilon^{\frac{3}{2}})) &O(\varepsilon) &O(\varepsilon^{\frac{1}{2}}) \\
O(\varepsilon) &O(\varepsilon^{\frac{1}{2}}) & 1
\end{pmatrix},
\end{equation}
where $\varepsilon$ is O$(\frac{m_c}{m_t})\approx 0.007$. If we restrict to the problem to a $2\times 2$ matrix with
mass eigenvalues of order $m_t$ and $m_c$,   the matrix must take a form so that the 
determinant turns out to be O($\varepsilon$): 
\dis{
m_t\begin{pmatrix}
O(\varepsilon) &O(\varepsilon^{1/2})   \\ 
O(\varepsilon^{1/2})   & 1
\end{pmatrix}.
}
In this case, the CKM matrix is close to a diagonal matrix \cite{Weinberg77}, which  basically results  
from  the huge mass hierarchy between $M_t$ and $M_c$. Extending this form to a $3\times 3$ 
matrix is  straightforward \cite{Fritzsch79}.

\vskip 0.5cm
\noindent{\bf 2. The PMNS Matrix}\vskip 0.5cm 

The PMNS matrix elements are
\dis{
{\rm PMNS}~({\rm component}\,ij):~&\frac{g}{\sqrt2}\, \bar{\nu}^{(0)}_{iL}V^{(\rm PMNS)}_{ij} \, 
\gamma^\mu   \, {e} _{jL}\, W^+_\mu\,,
\label{eq:PMNSij}
}
where $\nu^{(0)}_i$ and $e_j$ are the weak eigenstates. In addition, $e_{j}$ is also the mass eigenstate. 
For neutrino masses,   no counterpart corresponding to Eq. (\ref{eq:TopBasic}) exists instead, we 
obtain them from Weinberg's dimension 5 L-violating operator:
\dis{
 m_i \, \tilde{\nu}_{iL}C^{-1}\nu_{iL}.
}
 The diagonal masses of neutrinos are, for  the $(ll)$ component, 
\dis{
{\rm Diagonal}~(ll):&~ m_{\nu_1}\, \bar{\nu}_{1R}\nu_{1L}+m_{\nu_2}\, \bar{{\nu}}_{2R}\nu_{2L}+m_3\, 
\bar{\nu}_{3R}\nu_{3L}+{\rm h.c.}\\
\to &~ \frac12\left(m_{\nu_1}\, {\nu}_{1L}^TC^{-1}\nu_{1L}+m_{\nu_2}\,  {\nu}_{2L}^TC^{-1}\nu_{2L}+m_3\, 
 {\nu}_{3L}^TC^{-1}\nu_{3L}\right) \\
=\frac12\Big[m_{\nu_1}& \tilde{\nu}^{(0)}_{iL} \,(U^{(\nu)})_{1i} C^{-1} (U^{(\nu)\dagger})_{1j}\nu^{(0)}_{jL}+ 
m_{\nu_2}\tilde{\nu}^{(0)}_{iL} \,(U^{(\nu)})_{2i} C^{-1} (U^{(\nu)\dagger})_{2j}\nu^{(0)}_{jL}\\
&+m_{\nu_3}\tilde{\nu}^{(0)}_{iL} \,(U^{(\nu)})_{3i} C^{-1} (U^{(\nu)\dagger})_{3j}\nu^{(0)}_{jL}\Big]  ,
}
where in the second line we wrote the low-energy effective lagrangian in the SM in terms of L-handed fields as
\dis{
\nu_i = (U^{(\nu)\dagger})_{i\alpha}\nu^{(0)}_\alpha.
}
If a hierarchy of masses exists, we can follow the case of the CKM hierarchy and obtain 
\dis{
m_3\begin{pmatrix}
O(\varepsilon') &O(\sqrt{\varepsilon'})   \\ 
O(\sqrt{\varepsilon'})   & 1
\end{pmatrix}.
}
In the inverted hierarchy, the masses are almost degenerate, so the PMNS matrix need not be
near the identity matrix. From the data for the normal hierarchy \cite{NuFIT18}, we obtain $\Delta m^2\simeq 0.24
\times 10^{-4} \eV^2$ and $0.76\times 10^{-6} \eV^2$, {\it i.e.},   $m_3\simeq 0.5\times 10^{-2}$, and $m_2
\simeq 0.9\times 10^{-3}$, 
where $\varepsilon'$ is O($0.18$)=O($(0.4)^2$). Therefore,   the PMNS matrix being close 
to the identity matrix is not possible. This is because no strong hierarchy of neutrino masses exists. 
With a normal hierarchy of neutrino masses, for example, putting $M(\nu_1) \simeq 0$, the squared mass 
differences $\triangle_{12}^2 = 7.55\times 10^{-5} \eV^2$ and 
$\triangle_{23}^2 = 2.31\times 10^{-3} \eV^2$, which are measured \cite{NuFIT18}, suggest that $M(\nu_3) 
= 0.048 \eV$
and $M(\nu_2) = 8.7 \times 10^{-3} \eV$,  hence, the lepton hierarchy is $M(\nu_3)/M(\nu_2) = 5.5$.

In the quark sector, the top and the charm quarks have masses $M(t) = 173 \gev$ and
$M(c)= 1.2 \gev$, so the quark hierarchy is $M(t)/M(c) = 144.2$. Thus, the quark hierarchy
is much stronger than the lepton hierarchy.

\section{Flavor symmetry}\label{sec:Flavor}

If the PMNS matrix is not close to identity, the next question to ask is, ``How close are the PMNS-matrix elements?'' 
One may consider the mass matrix first, but experimentally presented ones are on the CC interactions. If a symmetry 
is the basis for large mixing angles in the PMNS matrix, the first one to consider is the permutation symmetry because 
permutation requires somethings to be identical. The simplest case $S_2$ has representations of singlets only; 
hence, it is not suitable for relating mixing angles. The permutation of three objects $S_3$ has a doublet representation, 
which may equate two entries among the matrix elements.  
 
 Let us consider a matrix with one zero,
 \dis{
 V^{(0)}\sim \begin{pmatrix}
 \times & \times &  \times \\
 \times & \times & \times \\
0 & \times & \times
 \end{pmatrix}
 }
 such that the CC in the leptonic sector is, viz. Eq. (\ref{eq:PMNSij}),
 \dis{
 &\overline{\nu}^{(0)}_LV^{(0)}\gamma^\mu e_L\\
 &=\overline{\nu}^{(0)}_L   U^{(\nu) \dagger} U^{(\nu)}V^{(0)}\gamma^\mu e_L\,, \label{eq:leptonCC}
 }
 where $e_L$ is a column vector of mass eigenstates, and
 \dis{
 \nu^{(0)}_L=U^{(\nu)}\nu .
 }
 Noting Eq. (\ref{eq:VandU}), $\nu^{(0)}_3= U^{(\nu)}_{3j}\nu_j$,  $ \nu^{(0)}_3$ is composed of a representation {\bf 2} of the 
 permutation symmetry $S_3$.  Negligible couplings occur between  {\bf 1} and {\bf 2} in $\nu^{(0)}_3$.
The PMNS matrix of the form in Eq. (\ref{eq:PMNSij}) is $S_3$ symmetric, and it is bimaximal; thus,
 \dis{
 V^{(\rm PMNS)}\sim \begin{pmatrix}
 \times & \times &   \times \\
 \times & \times & \times \\
0 &\pm \frac{ 1}{\sqrt2} & \frac{1}{\sqrt2}
 \end{pmatrix},
 }
 which may be obtained from the permutation symmetry $S_3$ \cite{Perkins95}.

By the same token, let us look for a permutation symmetry with a triplet representation {\bf 3}. It is present in
the permutation symmetry $S_4$.   In the PMNS matrix, can one associate 4 entries such that it contains a  triplet, for example,
 \dis{
 V^{(\rm try)}\sim \begin{pmatrix}
 \times & \times &   \times \\
 p & q & r \\
\circledS &\times & \times
 \end{pmatrix},
 }
  where we have declared that   a permutation symmetry of  $p,q,r,$  and $\circledS$ exists. We use $\circledS$ in anticipation of setting this entry to zero at the end. A triplet is assigned in  the second row of $V_{iI}$, {\it i.e.}, at $i=2$. Twenty-four elements of permutations of $p,q,r,\circledS$ are
 \dis{
 123\circledS,~231\circledS,~312\circledS,\\
12\circledS 3,~23 \circledS 1,~31 \circledS 2,\\
1 \circledS 23,~2  \circledS 31,~3  \circledS 12,\\
 \circledS 123,~ \circledS 231,~ \circledS 312, \label{eq:tvplus}
 }
 and
 \dis{
 213\circledS,~132\circledS,~321\circledS,\\
21\circledS 3,~13 \circledS 2,~32 \circledS 1,\\
2 \circledS 13,~1  \circledS 32,~3  \circledS 21,\\
 \circledS 213,~ \circledS 132,~ \circledS 321.\label{eq:tvminus}
 } 
 
Equation (\ref{eq:tvplus}) represents  cyclic permutations while Eq. (\ref{eq:tvminus}) represents  anti-cyclic 
permutations, which is identical to the scheme that neglects Eq. (\ref{eq:tvminus}) and allows plus and minus values 
of $\circledS$ in Eq.  (\ref{eq:tvplus}).
Depending on a nonzero value of $\circledS$, all 24 elements in Eq. (\ref{eq:tvplus}) and Eq. (\ref{eq:tvminus}) 
form the symmetric elements $S_4$. The four elements $p,q,r,$ and $\circledS$ must be the same. When 
only Eq. (\ref{eq:tvplus}) is considered, the same conclusion is drawn. For $p=q=r=\Delta$ while $\circledS$ takes two values $\pm\Delta$.  
 Now, suppose $p=q=r=\frac{1}{\sqrt3}$ and $\circledS=0$; then, two values of $\circledS$ collapse to one, and  Eq. (\ref{eq:tvplus}) has only  twelve elements. Thus,   we argue that the discrete group of twelve elements, 
 $A_4$ \cite{Ma02}, will lead to a PMNS matrix of the form
 \dis{
V\sim  \begin{pmatrix}
 \times &\times& \times\\[0.5em]
\frac{1}{\sqrt3}& \frac{1}{\sqrt3}&\frac{1}{\sqrt3} \\[0.7em]
0 & \sin\alpha & \cos\alpha
\end{pmatrix}. \label{eq:PMNSForm}
 }
For a tri-bimaximal form, we require $\alpha=\pm 45^{\rm o}$, for which $V_{iI}$ has an additional permutation 
 symmetry: $V_{i2}\leftrightarrow V_{i3}$ for  $i=3$. Note that Eq. (\ref{eq:PMNSForm}) uses the weak 
 eigenstates of neutrinos with superscripts $^{(0)}$.
 
  We can present our arguments in terms of the $A_4$ representations. $A_4$ has three singlets 
 $\one,\one',$ and $\one''$ and one triplet $\three$.  How then three singlets are given is obtained from the 
 decomposition of $S_4$ representations into those of $A_4$, as shown in Table 1. The singlet 
 representations are $\one=p+q+r+\circledS,\one'=p+q-r-\circledS$ and $\one''=p-q+r-\circledS$.   
 The three numbers in the first row of Eq. (\ref{eq:PMNSForm}) correspond to three singlets.  The two numbers in 
 the 3rd row  correspond to two singlets $\one'$ and $\one''$; hence, $\alpha$ need not be 45$^{\rm o}$. 
Hence, Eq. (\ref{eq:PMNSForm}) can be written in terms of $A_4$ representations as
 \dis{
V\sim  \begin{pmatrix}
 \one &\one'& \one''\\[0.5em]
 & \three^T&  \\[0.7em]
0 & \one' & \one''
\end{pmatrix}. \label{eq:PMNSfromS4}
 }

 \begin{table}[!h] 
\caption{\label{tab:S4A4} Branching of $S_4$ representations $\one, \one',\two, \three$ and $\three'$ into 
the $A_4$ and the $S_3$ representations \cite{KobaBk12}.}
\begin{tabular}{c|c|c}
\hline
 &\\[-1.3em]  
 $S_4$~& $A_4$ & $S_3$\\ \hline
   $\one$~ & $\one$&$\one$  \\[0.3em]
  $\one'$ &$\one$&$\one'$  \\[0.3em]
  $\two$~ &~~$\one'\oplus \one''$~~&$\two$  \\[0.3em]
  $\three$~ &$\three$&$\one\oplus \two$  \\[0.3em]
  ~$\three'$~~&$\three$&~$\one'\oplus \two$~\\[0.3em]
  \hline
  \end{tabular}
\end{table}

\section{Model for Neutrino Masses}\label{sec:Model} 
 
To realize $A_4$ symmetry, we assign the Yukawa couplings such that the flavor indices of $i$ respect the   
requirements discussed in Section \ref{sec:Flavor}. The L-violating BSM singlets $N_i$ can be three, but here 
we show an example with just one singlet $N$: ${\cal L}_{\Delta L}=\frac12m_N N^2$. The effects of other 
singlets is effectively adjusted by the coefficients of the Weinberg operator,
\dis{
 \frac{v_u^2}{m_N} \, \tilde{\nu}^{(0)}  _{iL}C^{-1}(h_{ij})\nu^{(0)}  _{jL},\label{eq:WeinNu}
} 
where ($h_{ij}$) denotes the Yukawa matrix.  
In Eq. (\ref{eq:WeinNu}), the hierarchy of $h_{ij}$ is determined by physics above the electroweak scale, in particular by the VEVs of the SM singlet Higgs fields rendering the heavy neutrinos mass. For example, if the heavy neutrinos are completely democratic, the heavy neutrinos have a hierarchy $M_3\gg M_1=M_2\approx 0$, approximately leading to $h_{11}=h_{22}\gg h_{33}$. With the $A_4$ symmetry, the heavy neutrino hierarchy is determined by the representation property of the singlet Higgs. This feature is discussed in \cite{FKmore}.
Because we introduce just one Higgs doublet $H_u$ giving mass 
to neutrinos, the $A_4$ symmetry is counted only by the neutrinos.
 The weak eigenstate neutrinos of Eq. (\ref{eq:WeinNu}) are related to the mass eigenstate neutrinos by, 
 neglecting phases of a tri-bimaximal $U$, 
 \dis{
  \nu^{(0)}_{\theta\phi}=U(\theta,\phi)\nu  \sim \begin{pmatrix}
\cos\theta, ~\sin\theta\cos\phi , ~ \sin\theta\sin\phi\\[0.5em]
 {\bf 3}^T \\[0.7em]
0, ~\frac{\pm 1}{2},~\frac{ 1}{2}
\end{pmatrix}\begin{pmatrix}\nu_e\\[0.5em] \nu_\mu \\[0.7em] \nu_\tau\end{pmatrix} 
= \begin{pmatrix}\cos\theta\,\nu_e  + \sin\theta\cos\phi\,\nu_\mu+ \sin\theta\sin\phi\,\nu_\tau  \\[0.5em] 
\frac{1}{\sqrt3}\nu_e + \frac{1}{\sqrt3}\nu_\mu +\frac{1}{\sqrt3}\nu_\tau  \\[0.7em] \frac{\pm1}{\sqrt2}\nu_\mu 
+ \frac{1}{\sqrt2}\nu_\tau \end{pmatrix},  \label{eq:MassNu}
 }
and Eq. (\ref{eq:WeinNu}) expressed in terms of the mass eigenstates is, for $\cos\theta=
\frac{\pm\sqrt2}{\sqrt3}$ and $\phi=45^{\rm o}$,
 \dis{
 h\,\frac{v_u^2}{m_N} &\Big( \left[\frac{\pm\sqrt2}{\sqrt3}\,\tilde{\nu}_e  + \frac{1}{\sqrt6}\,\tilde{\nu}_\mu
 + \frac{1}{\sqrt6}\,\tilde{\nu}_\tau \right]C^{-1}  \left[\frac{\pm\sqrt2}{\sqrt3}\,\nu_e  +\frac{1}{\sqrt6}\,\nu_\mu
 + \frac{1}{\sqrt6}\,\nu_\tau  \right]\\[0.5em]
 &+\left[\frac{1}{\sqrt3}\tilde{\nu}_e + \frac{1}{\sqrt3}\tilde{\nu}_\mu  +\frac{1}{\sqrt3}\tilde{\nu}_\tau  \right]
 C^{-1}\left[\frac{1}{\sqrt3}\nu_e  + \frac{1}{\sqrt3}\nu_\mu +\frac{1}{\sqrt3}\nu_\tau \right]\\[0.7em]
 &+\left[  \frac{\pm1}{\sqrt2}\tilde{\nu}_\mu + \frac{1}{\sqrt2}\tilde{\nu}_\tau  \right]C^{-1}\left[ \frac{\pm1}{\sqrt2}
 \nu_\mu + \frac{1}{\sqrt2}\nu_\tau \right]\Big).\label{eq:Wein0s}
 }
 Certainly, Eq. (\ref{eq:Wein0s}) looks a bit more complicated than Eq. (\ref{eq:WeinNu}). 
 
 \bigskip
 
Generalizing   the forms $h_{ij}$ and $U$ is straightforward  if the the three heavy neutrino masses are different 
 and the three CP phases are included. Nevertheless, if the (31) element of $U$ remains   zero as in Eq. 
 (\ref{eq:Wein0s}), then  the Dirac CP phase does not appear \cite{KimSeo12}. For the Dirac 
 CP phase to have an effect,  a  term(s) violating the tri-bimaximal form, notably in the (31) element of $U(\theta,\phi)$, must exist.
 
A model construction is to declare the  weak eigenstates of neutrinos with superscripts $^{(0)}$ to be the triplet 
representation under the dihedral group $A_4$ and $V^{(0)}$ in Eq. (\ref{eq:leptonCC}) to be Eq. (\ref{eq:PMNSfromS4}).  
Of course, we work in bases where the $T_3=-\frac12$ components in the quark and lepton sectors are 
mass eigenstates. At this stage,  all quarks states can be taken as singlets under $A_4$. In GUTs, the tensor 
product  ${\bf 3}\times \three= 2\cdot \three \oplus \one\oplus\one'\oplus \one''$ should be considered to 
declare the $A_4$ properties of the quark representations, which will be discussed in the future \cite{FKmore}. 

\section{Conclusion}\label{sec:Conclusion} 

Our letter suggests a general approach to study the connection between the lepton and the
quark parameters precisely by exploiting the difference between the two sectors. We have compared the 
 mixing angles and the masses pertinent to the quark and the lepton
mixing matrices. We have shown that the two matrices possess intriguing hierarchies 
anticorrelated between angles and masses and hope this observation will lead to
further progress.

 \vskip 0.5cm
\centerline{ACKNOWLEDGMENT }
This work is supported in part  by a   National Research Foundation (NRF) grant from Korea (NRF-2018R1A2A3074631).


\end{document}